\documentclass[fleqn]{annalen}
\usepackage{epsf,psfig}
\usepackage{amsmath,amssymb}
\newcommand{\figwidth}{0.85\columnwidth}
\begin{document}
\newcommand{\volume}{11}              
\newcommand{\xyear}{2002}            
\newcommand{\issue}{12}               
\newcommand{\recdate}{11 November 2002}    
\newcommand{\revdate}{dd.mm.yyyy}    
\newcommand{\revnum}{0}              
\newcommand{\accdate}{dd November 2002}    
\newcommand{\coeditor}{B. Kramer}           
\newcommand{\firstpage}{1}           
\newcommand{\lastpage}{15}            
\setcounter{page}{\firstpage}        
\newcommand{\keywords}{quantum spin systems, disordered wires}
\newcommand{\PACS}{75.10.Jm, 75.40.Ng}
\newcommand{\shorttitle}{C. Schuster and U. Eckern, Quantum spin 
chains with defects} 
\title{Quantum 
spin chains with various defects}
\author{Cosima Schuster$^*$ and Ulrich Eckern}
\newcommand{\address}
{Institut f\"ur Physik, Universit\"at Augsburg, 86135 Augsburg, Germany}
\newcommand{\email}{$^*$ Corresponding author: cosima.schuster@physik.uni-augsburg.de}
\maketitle
\begin{abstract}
Using the density matrix renormalization group
(DMRG) method, we study the quantum coherence in one-dimensional
disordered spin chains and Fermi systems.
 We consider in detail spinless fermions on a ring,
and compare the influence of several kinds of impurities
in a gapless and a dimerized, gapped system.
In the translation-invariant system a  so-called
site-impurity, which can be realized by a local potential  or a
modification
of one link, 
increases for repulsive interaction, and decreases for attractive
interaction, upon renormalization.
The weakening of two neighbouring bonds, which is a realization of
a so-called bond-impurity, on the other hand,
is  healed for
repulsive interaction, but enhanced for intermediate attractive interactions.
This leads to a strong suppression
of the quantum coherence measured by the phase sensitivity, but not to localization.
Adding  a local distortion to a dimerized system, we find
that even the presence of a single site-impurity increases  the
metallic region found in the dimerized model.
For a strong dimerization and  a high barrier, an
additional sharp maximum, is seen in
the phase sensitivity as a function of interaction, already for systems
with about 100 sites.
A bond-impurity in the dimerized system
also opens a small metallic window in the otherwise isolating regime.
\end{abstract}
\section{Introduction}
Recent experiments have led to a renewed theoretical interest in 
disordered spin-Peierls systems. For example, 
the doped one-dimensional Heisenberg system 
Cu$_{1-x}$Zn$_x$GeO$_3$ \cite{Exp97}
shows two subsequent magnetic transitions,
the spin-Peierls transition at $T=14$ K followed by a transition 
 to an antiferromagnetic
ground-state at $T_{\rm N}=5$ K, while in
Cu$_{1-x}$Mg$_x$GeO$_3$, a reentrant spin-Peierls phase
for $x>x_c$ \cite{exp00} is observed.
In fact, the  general question of  the effect of various types of
impurities in pure \cite{Eggert92,Qin,Meden98,Schmi95,Kluem98,Fulde}
or gapped 
spin systems \cite{Bhatt,Fabrizio97,Most97} has been studied 
intensively during the last years. Nevertheless,
the interplay between interaction, disorder, and periodic distortions
is still a challenging problem. 
The model of spinless fermions, which is considered here
in detail, is equivalent to the anisotropic Heisenberg model.
It describes certain aspects of magnetic and electronic systems, 
and the phase diagram of this ``simplest" interacting fermion 
model is surprisingly rich. On the basis of this model,
the aim of this work is thus to achieve a better understanding of the 
ground-state properties of disordered spin- and interacting Fermi-systems,
especially clarifying the role of the 
interaction, which may enhance or decrease
the localization due to the random and periodic
perturbations. In particular, we will
introduce and compare the effects of various kinds of impurities.

In this context,  Eggert and Affleck \cite{Eggert92} pointed out
that two ``classes" of impurities exist, which differ in
their effect on the local symmetry of the system.
  So-called ``site-impurities" violate the site parity
by affecting one site or bond,
whereas ``bond-impurities", which modify two neighbouring bonds, violate the
bond  symmetry but respect the site parity.
Bearing this in mind, we use the name site- or bond-impurity.
Note that this naming is different from the one used in \cite{Qin} and
\cite{Meden98}, where an impurity on a single site is called  site-impurity,
 and an impurity on a bond is called bond-impurity.
Another type, ``transparent" impurities, which we do not consider here, 
can be constructed using the Bethe-Ansatz \cite{Schmi95}:
Those defects are similar to bond-impurities, but include 
also a coupling between the
next-nearest neighbours around the impurity. Transparent impurities
have also been studied extensively with analytic and numerical techniques
during the last years, see for example \cite{Kluem98}
and \cite{Fulde}.
As a result of Eggert and Affleck,
a site-impurity is relevant in the sense that it can break up a closed ring,
whereas a bond-impurity is irrelevant, i.~e.\ the defect is healed 
at low energies.

Gapped spin systems containing irrelevant impurities,
i.~e.\ impurities which do not close the gap in the system,
are supposed to be equivalent to free Dirac-fermions with random mass, 
a model which has been widely investigated in 
one dimension \cite{Balents97,Gogolin97} in the context of doped
spin-Peierls or spin-ladder systems and two dimensions \cite{Ziegler} 
in the context of the Quantum Hall Effect. 
However, the transformation to Dirac fermions is valid only for special
points in the parameter space, which correspond to the XY model (free fermions)
for the dimerized system,
or the isotropic XXX-Heisenberg antiferromagnet for the ladder model.

The numerical density matrix renormalization group \cite{White92} is
a quasi-exact numerical method to determine the ground state properties,
in particular the ground-state energy,
of long one-dimensional (non-integrable) systems with reasonable accuracy.
The bosonization technique \cite{Haldane81} can be used with advantage
to interpret the numerical data.

In the following, we introduce the models and 
the impurity types studied by us.
Using the bosonization technique and the Luttinger description,
we then identify the leading non-linear operators,
and classify the impurities according to them.
The results for single defects
 are presented in  Sec. \ref{sec3}. 
Concerning the dimerized system, we concentrate in our
discussion on strong site- and bond-impurities.
The results are discussed in  Sec. \ref{sec4}, followed by a summary in
 Sec. \ref{sec5}.

\section{Magnetic chains and spinless fermions}
We begin by presenting  
the generic spin model, and then describe
the equivalent fermionic model. Then we briefly introduce
the Luttinger description,  
which is  useful for  discussing the relevant operators
which lead to insulating behaviour. 
\subsection{The Heisenberg spin chain}
As a starting point for the study of a general disordered spin-Peierls system 
in one 
dimension, we consider an anisotropic Heisenberg chain,
given by the  XXZ model, with a dimerized interaction: 
\begin{equation} 
\label{spins} 
H_{\rm spin}= -\sum^{N}_{i=1} 
J_i (u) \left( \sigma^{x}_{i} \sigma^{x}_{i+1} + 
\sigma^{y}_{i} \sigma^{y}_{i+1} 
   + \Delta \sigma^{z}_{i} \sigma^{z}_{i+1}\right) 
 + N\frac{K}{2}u^2, 
\end{equation} 
where the dimerization in the Peierls state, $u$, enters the coupling
constant according to
$J_i (u) =J [1+(-)^i u]$. 
For the clean XXZ model, i.~e.\ for $u=0$, and for zero total
magnetisation, $M=\sum_i\langle\sigma^z_i \rangle =0$, 
one finds three
phases:
a ferromagnetic phase for $\Delta \geq1$, separated by a first-order transition
from 
a gapless phase for $-1\leq\Delta < 1$, whose low lying excitations 
are given by those of a Luttinger liquid;
and an antiferromagnetic phase for 
$\Delta < -1$. The transition from the Luttinger to the antiferromagnetic
phase is of 
Berezinskii-Kosterlitz-Thouless type.
Adding the dimerization, $u$, the system becomes localized by forming
spin-singlets on neighbouring sites  for $\Delta<0$, i.e. 
for antiferromagnetic coupling.
An excitation
gap opens due to the usual Peierls mechanism \cite{Peierls}. Leaving aside
the question whether a finite $u$ can be stabilized, we note
that the dimerization
is already relevant for $\Delta<\sqrt{2}/2$ 
 \cite{koh81,alc95},
and the ground-state wave
function is localized.
The interaction-dimerization phase diagram was determined 
in \cite{EPJB}.
 
Impurities may be realized in three different ways.
First, local magnetic fields may be present, resulting for example 
from magnetic impurities near the chain,
which couple directly to the $\sigma^z$-component at a specific site.
We call this kind of impurity a site-impurity. 
Second, the coupling between the spins may be modified locally by a factor 
$(1-\delta)$, for example
by substituting Sr with La or Ge with Si
in a spin chain compound like La$_x$Sr$_{2-x}$CuO$_3$ or 
CuGe$_{1-x}$Si$_x$O$_3$.
As only one link is concerned this is similar to a 
local field;
nevertheless we call it link-impurity in the following.
The third possibility is to modify
both bonds
to the left and to the right of the impurity site by an equal amount, $(1-b)$:
This situation, which is called bond-impurity,
can be realized by doping magnetic impurities with spin 1/2 into a chain,
e.~g.\ substituting V by Nb. (Nevertheless, in some cases it is 
found that the free Nb-electron 
 -- with the spin --
moves to the Vanadium site, i.~e.\ a Nb$^{\rm 5+}$- and a V$^{\rm 3+}$-site 
is established instead of
Nb$^{\rm 4+}$ and V$^{\rm 4+}$.) 
Completing the Hamiltonian \eqref{spins} in this way, we can write
\begin{equation}
\label{spins-imp}
H_{\rm spin}= -\sum^{N}_{i=1}
J_i(u,\delta,b)  \left( \sigma^{x}_{i} \sigma^{x}_{i+1} +
\sigma^{y}_{i} \sigma^{y}_{i+1}
   + \Delta \sigma^{z}_{i} \sigma^{z}_{i+1}\right)
 - \sum^{N}_{i=1} h_i\sigma^z_i\;,
\end{equation}
with $J_i(u,\delta,b)=J_i(u)[1-\delta_{i,m}\delta-
          (\delta_{i,m}+\delta_{i,m+1})b]$, where $m$ is a fixed site.

\subsection{The fermionic model}
The corresponding fermionic model is obtained 
via the Jordan-Wigner transformation.
In the result we change the notation, $J\to t$, $J\Delta \to -V/2$, 
and $\epsilon_m \to - 2h_m$, and neglect  
constant energy shifts, like  $\sum_i h_i$ and $\Delta (2N_f-N/2)$,
where $N_f$ is the number of fermions: 
\begin{eqnarray}
\label{fermi}
 H_{\rm fermion} &=&  - \sum_{i} t_i 
\left( c^+_ic^{\hphantom +}_{i+1} +c^+_{i+1}c^{\hphantom +}_{i} \right) 
      + \sum_i  V_i   n_i n_{i+1} +
\sum_i \epsilon_in_i \\
&&-\frac{V}{2}\left[\delta (n_m+n_{m+1})+
b(n_{m-1}+2n_m+n_{m+1})\right],\label{secondline}
\end{eqnarray}
where $t_i (u) = t [1+(-1)^iu_t-\delta_{i,m}\delta_t-(\delta_{i,m}+\delta_{i,m+1})b_t]$, and
$V_i (u) = V [1+(-1)^iu_V-\delta_{i,m}\delta_V-(\delta_{i,m}+\delta_{i,m+1})b_V]$.
Furthermore, we have assumed
that the coupling to (static) phonons and to the impurities
can be varied independently in
both the hopping and the interaction, by introducing six parameter
($u_t, u_V, \delta_t, \delta_V, b_t, b_V$) instead of three 
($u, \delta, b$). 
The particle density is $n_0=N_f/N$, and we restrict ourselves
to half filling; we set $t=1$ in some of the
formulas below.
As shown in  \cite{Diss}, $u_t$ is the main contribution arising from
the dimerization, whereas $u_V$ modifies the results only
quantitatively but not qualitatively; thus we assume $u_V=0$.
For simplicity, we likewise neglect  \eqref{secondline}
to avoid the mixing -- in the fermionic picture --
of bond- and site-impurity.
In the case of the link-impurity,
it was shown by Meden et al. \cite{Meden98} that  the weakening of
one bond
in the spin model is equivalent to an impurity given by 
$\delta_t$ and $\delta_V$, i.~e.\ that the contribution \eqref{secondline}  
is of lesser importance.
We assume that the same holds for the bond-impurity. 
The in this way simplified Hamiltonian is then given by
\begin{eqnarray}
\label{fermisim}
H_{\rm fermion} &=&  - \sum_{i} t[1+(-1)^iu_t-\delta_{i,m}\delta_t-(\delta_{i,m}+\delta_{i,m+1})b_t]
\left( c^+_ic^{\hphantom +}_{i+1} +c^+_{i+1}c^{\hphantom +}_{i} \right)
\nonumber \\&&      
+ \sum_i  V[1-\delta_{i,m}\delta_V-(\delta_{i,m}+\delta_{i,m+1})b_V]
+\sum_i \epsilon_in_i.
\end{eqnarray}
Thus the link-impurity can occur as a pure hopping-impurity,
$\delta_t\neq 0$ and $\delta_V=0$, or a generic link-impurity,
$\delta_t \neq 0$ and $\delta_V\neq 0$.
Setting  $\delta_t=0$, we can also 
realize an interaction-impurity by varying $\delta_V$.
In the following we consider the Hamiltonian given by Eq. \eqref{fermisim} 
only.

\subsection{Bosonization and Luttinger description}
For the interpretation of the numerical data, we use the fact that 
in the gapless phase the system can 
be described by a Luttinger liquid,
and that the distortions can be considered as a perturbation.
In the bosonized form \cite{Haldane81}, the  Hamiltonian can be written 
for the ``clean'' model as follows:
\begin{equation}
\label{boson}
H_{\rm boson} = \int \frac{{\rm d}x}{2\pi}\ \left\{\frac{v}{g}
\left[\partial_x\varphi(x,t)\right]^2 +
vg\left[\pi\Pi_\varphi(x,t)\right]^2 \right\},
\end{equation}
where $\Pi_\varphi$ is the momentum density conjugate to $\varphi(x,t)$,
i.~e.\ it corresponds to
$\Pi_\varphi=\partial_t\varphi/(vg)$ in the (Euclidean) path integral
formulation of the theory. It is
also related to the conjugate phase variable $\theta$ through
$\Pi_\varphi=\partial_x\theta(x)/\pi$.
The velocity $v$ of the bosonic excitations is given by 
$v={[\pi t\sin(2\eta)]/( \pi -2\eta)}$, and the interaction constant is 
$g=\pi / 4\eta$,
where $\eta$ parameterizes the interaction according to
$V=-2t\cos(2\eta)$. 
The density (operator) is given by
\begin{equation}\label{dens}
n(x)=n_0+\frac{\partial\varphi}{\pi\partial x}+
\frac{k_F}{\pi}\cos{[2k_Fx+2\varphi(x)]}.
\end{equation}
We chose the above standard representation because then the
order of the scattering process is  directly seen in the 
non-linear terms.
Non-linear -- global and local -- contributions arise  from the 
 dimerization
and the interaction as well as from  the impurities.
\begin{itemize}
\item Dimerization  causes  a $2k_F$-scattering process of the fermions,
\begin{equation}\label{hamu}
H_{u}\propto 2u\int {\rm d}x \; \sin[{2\varphi(x)}].
\end{equation}
\item While the fermion-fermion scattering with
$q\approx 0$ and $q\approx 2k_F$ is absorbed in the Luttinger-parameter $g$,
the  $4k_F$-scattering leads to 
\begin{equation}\label{hamV}
H_{V}\propto V\int \text{d}x \; \cos{[4\varphi(x)+(4k_Fx-G)x]},
\end{equation}
where $G$ is the reciprocal lattice vector.
This term causes the transition to the CDW-state for $V=2$ ($g=1/2$).
\item 
Backscattering arises from a local potential (say at $x=0$).
Since the potential couples directly to the density, Eq. \eqref{dens},
we find in lowest order in the impurity strength
\begin{equation}\label{hame}
H_{\epsilon}=\epsilon_0n_0\propto  \epsilon_0n_{2k_F}(x=0) \propto \epsilon_0 \cos[2\varphi(0)].
 \end{equation}
A strong potential in an
infinitely long chain is equivalent to a weak link \cite{Kane92}:
The diagonalization of the above Hamiltonian, $H_{\rm fermion}+H_{\epsilon}$,
leads, in the case of a strong potential between two semi-infinite chains
($R$: right chain, $L$: left chain),
to the contribution
\begin{equation}\label{hames}
H_{\epsilon}\propto \frac{c_R^+(0)c_L^{}(0)}{\epsilon_0}
\propto \frac{1}{\epsilon_0} \cos[2\theta(0)].
 \end{equation}
\item A change in the hopping 
is also considered in the weak, $\delta_t \to 0$,
and strong, $\delta_t \to 1$, limit.
Following the analysis for the (weak) periodic \cite{Diss} or random potential \cite{Doty},
we would assume that 
the transition from a site-centered potential to a bond-centered 
potential causes only a phase shift of $\pi/2$ in the non-linear term,
i.~e.\ a shift from a cosine to a sine. 
The weak link thus should correspond to a high barrier, 
see again \cite{Kane92}, i.~e.
\begin{eqnarray}
H_{t}&\propto& \delta_t \sin[2\varphi(0)] \; , \quad\quad
   \delta_t\to 0 \label{hamt1s}  \\
H_{t}&\propto& (1-\delta_t)\cos[2\theta(0)] \; , \quad\quad
   \delta_t\to 1 \label{hamt2s}
\end{eqnarray}
should be  appropriate descriptions.
However, if we consider  a modified link 
with $H_t\propto (1-\delta_t)(c_1^{+}c_0^{}+\rm{h.~c.})$,
we  find instead
\begin{eqnarray}
H_{t}&\propto& \delta_t\sin[\varphi(1)\sin[\varphi(0)]
 \label{hamt1}\\
H_{t}&\propto&(1-\delta_t) \cos[\theta(1)-\theta(0)]
\label{hamt2}
.
\end{eqnarray}
\item The local modification of the interaction, i.~e.\ the
interaction-impurity, leads to a $2k_F$-contribution too, in addition
to the  $4k_F$-term already present in Eq. \eqref{hamV}:
\begin{equation}\label{hamVl}
H_{\delta_V}\propto H_{t} 
\end{equation}
\item 
The modification of 
two links, in the limit of $b=b_t=b_V\to0$, does not cause $2k_F$-scattering
at half filling.
Thus, no non-linear contribution  to the Hamiltonian is found.
In the limit $b\to 1$, however, we find the following expression:
\begin{eqnarray}\label{hamb}
H_{b}&=&(1-b)(c^+_1c^{}_0+c^+_0c^{}_{-1}+{\rm h.c}) \nonumber \\
&\propto&  (1-b) \sin{\{[\theta(1)+\theta(-1)]/2-\theta(0)\}} 
\sin{\{[\theta(1)-\theta(-1)]/2\}} \nonumber \\ && \hphantom{(1-b)}\times\cos[\varphi(0)]
\sin\{[\varphi(1)+\varphi(-1)]/2\} \nonumber \\
&\propto& (1-b) 
\sin{(\Phi/2)}\sin{[\Theta(1)]}\cos[\varphi(0)]
\sin[\phi(1)].
\end{eqnarray}
In the last step we use 
$\theta(x)=\theta_0+\Phi x/L + \tilde\theta(x)$, see \cite{Haldane81}, where $\theta_0$ represents 
the zero-mode contribution,
$\Phi x/L$ the boundary condition -- see also the following section --,  
and $\tilde\theta(x)$ the excitations.
In addition, the abbreviations $\varphi(x)+\varphi(-x)=2\phi(x)$,
and $\theta(x)-\theta(0)=\Theta(x)$ are used. 
\end{itemize}

\subsection{The phase sensitivity}
We will use the phase sensitivity, i.~e.\ the reaction of the system to a 
change in the
boundary condition, to determine the localized-delocalized 
transition numerically for systems with finite
 size. 
We model the different boundary condition via a magnetic flux, 
penetrating the ring
of the spinless fermions.
The effect of the magnetic flux results in an additional   
phase of the hopping terms,
$t_j\to |t_j|{\rm e}^{i\Theta_j}$, with $-\pi<\Theta_j<\pi$.
The energy levels depend only on the total flux,
$\Phi={\rm arg}\left( \Pi_{j=1}^Nt_j\right)$.
In particular, we determine below
the energy difference between periodic ($c_N=c_0$, $\Phi=0$) and anti-periodic 
($c_N=-c_0$, $\Phi=\pi$) boundary conditions,
$\Delta E= (-)^{N_f}(E(0)-E(\pi))$. 
Here the factor $(-1)^{N_f}$ cancels the odd-even effects resulting from the 
change in the ground-state for odd versus even particle numbers. 
We recall that,
for a clean system, the ground-state energy and the finite-size corrections
can be obtained from the Bethe Ansatz \cite{Eckle87}.
At half filling (and for odd particle number), the result, 
in the Luttinger regime,
is given by \cite{Shastry90}
\begin{equation}\label{Bethe}
E_N(\Phi)-N\varepsilon_{\infty}=
-\frac{\pi v}{6 N}\left(1-3g\frac{\Phi^2}{\pi^2}\right),
\end{equation}
where $\varepsilon_{\infty}$ is the
energy density in the thermodynamic limit. Thus $N\Delta E=\pi v g/2$, 
independent of $N$, for the
metallic state. In an insulator, on the other hand, 
we find localized levels and 
the system cannot react to a twist in the boundary condition,
i.~e.\  $N\Delta
 E$ is expected to decrease with system size.
Considering the above model, in which 
the contributions of the local potential and the weakening of the 
neighbouring bonds
are taken into account separately, the 
localized levels do not split off the continuous spectrum \cite{Kleiner}, 
and the 
phase sensitivity is an appropriate
observable.

\section{Comparison of  site- and bond-impurities}\label{sec3}
\subsection{Local potential}
\begin{figure}
\begin{center}
\centerline{\psfig{figure=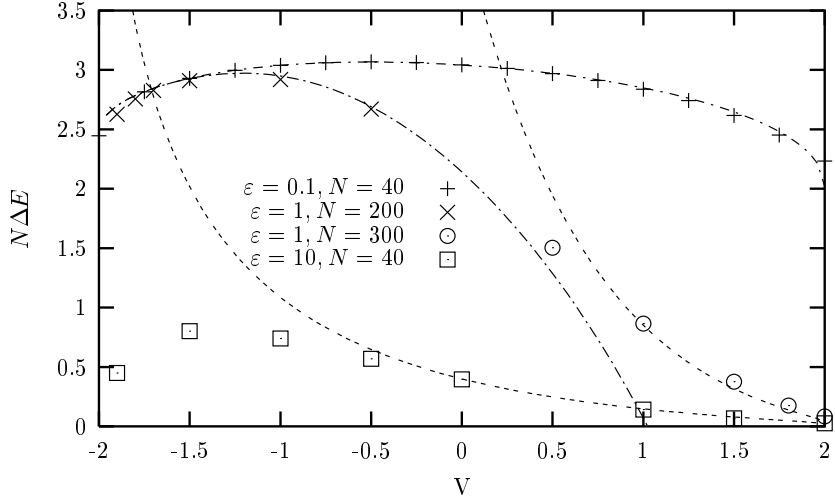,width=\figwidth}}
\end{center}
\caption{\label{fig1}Phase sensitivity as a function of 
interaction for a system with one
site-impurity. 
The dashed-dotted lines correspond to Eq. \eqref{eq13}, and the dashed lines to 
\eqref{eq15}.}
\end{figure}
A local magnetic field, which corresponds to a local potential, 
is the well-known and most studied example of a site-impurity.
In this case, the free motion of the fermions inside the ring is restricted
by the back-scattering  at the impurity
($\pm k_F\to \mp k_F$). 
As discussed by Kane and Fisher \cite{Kane92} the 
impurity 
becomes transparent for an attractive,
and completely reflective for a repulsive
interaction, according to the renormalization group equation
\begin{equation}\label{KF}
\frac{{\rm d}\epsilon_0}{{\rm d}l}=(1-g)\epsilon_0.
\end{equation}
In other words, the impurity strength scales either to zero or to infinity,
for $g>1$ and $g<1$, respectively.
The fact that an impurity scales to zero for repulsive interaction
is confirmed by the scaling of a weak link \cite{Kane92}.
For  strong $\epsilon_0$ the  effective hopping 
is given by $t_t=4t^2/\epsilon_0$.
The final result for the phase sensitivity is obtained by
 perturbation theory with respect to the defect strength, see 
\cite{Diss} and  \cite{SCH}, and is given by
\begin{equation}\label{eq13}
 N\Delta E = {\pi v g \over 2} - \epsilon_0 \left( {N\over N_0} \right)^{1-g}
\end{equation}
in the weak scattering limit,
and by
\begin{equation}\label{eq15}
N\Delta E = \frac{4t^2}{|\epsilon_0 |}\left( {N\over N_0} \right)^{1-1/g}
\end{equation}
in the strong impurity limit, where $N_0\approx 2$ is a cut-off corresponding 
to a momentum cut-off of the order of the Fermi momentum.
As a summary, we show numerical data in Fig. \ref{fig1}.
The most important result is that an intermediate defect strength scales
to zero
for attractive interaction, and to infinity
for repulsive interaction, rather than scaling
to an additional intermediate fixed point.
\subsection{One modified link}
\begin{figure}
\centerline{\psfig{figure=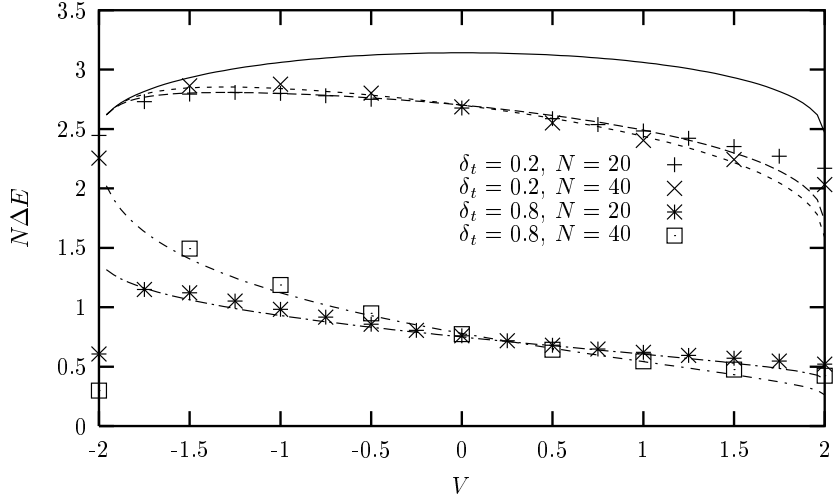,width=\figwidth}}
  \caption{\label{fig-tfit}Phase sensitivity as a function 
of interaction for a system with
a hopping impurity.
The straight line corresponds to the clean system, Eq. \eqref{Bethe}.
The dashed lines correspond to
Eq. \eqref{eq13t}, and the dashed-dotted lines to Eq. \eqref{eq15t}.}
\end{figure}

We begin by checking whether the above expressions for the
weak and strong potential, Eqs. \eqref{eq13} and
\eqref{eq15}, can be applied  to the case of a modified link,
as should be expected  from 
the treatment in \cite{Kane92} or \cite{Doty}.
We therefore calculate the phase sensitivity for a weakly modified link, 
$\delta_t=0.2$, and a weak link, $\delta_t=0.8$.
The results are shown in Fig. \ref{fig-tfit}.
However, the above expressions do not apply:
Instead of Eq. \eqref{eq13} we find for the weakly modified link 
\begin{equation}\label{eq13t}
 N\Delta E = {\pi v g \over 2} - 2\delta_t\left( {N\over N_0} \right)^{(1-g)/2}.
\end{equation}
This is actually the scaling according to Eq. \eqref{hamt1}.
Therefore, Eq. \eqref{hamt1s} does not  appropriately describe 
this impurity type. Similarly, for
the weak link, we do not recover Eq. \eqref{eq15},
but instead
\begin{equation}\label{eq15t}
 N\Delta E = 4(1-\delta_t)\left( {N\over 2N_0} \right)^{(1-1/g)/2}.
\end{equation}
In this case $N\to N/2$, as we connect two chains of length $N/2$.

\begin{figure}
{\psfig{figure=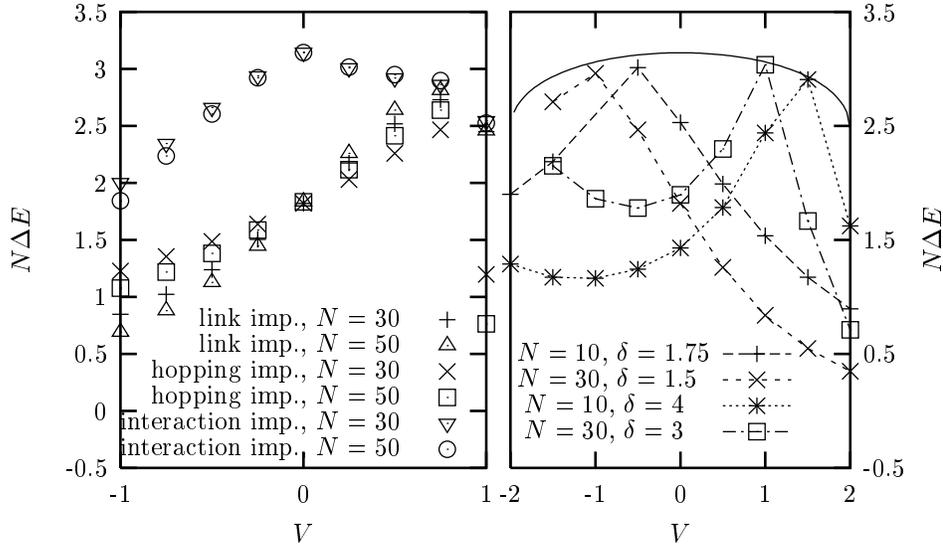,width=1.0\columnwidth}}
  \caption{\label{fig2} Left:
Phase sensitivity as a function of interaction for a system with one
site-impurity of type $t$, $tV$, and $V$.
Right: Details of the 
phase sensitivity as a function of interaction for a system with one
strong link-impurity. The straight line corresponds to the clean case, the 
other lines are connecting the data points.}
\end{figure}

Next we verify that the generic link-impurity
can be written as a sum of a hopping- and an interaction-impurity,
as the bosonized Hamiltonians, Eqs. \eqref{hamt1} and  \eqref{hamVl},
suggest. 
As shown on the l.~h.~s.\ of Fig. \ref{fig2},  where
the three cases: $\delta_t=\delta$ and $\delta_V=0$, $\delta_t=\delta_V$,
and $\delta_V=\delta$ and $\delta_t=0$, are compared, this conjecture
is indeed confirmed. 
The 
qualitative behaviour is the same for all defect types. 
By comparing two different system sizes,
we can fix the transition point to $V=0$
in all cases.

If the impurity is so strong that the sign of the
hopping or the interaction changes, $\delta >1$, 
interaction dependent
maxima occur, 
in addition,
for the generic link-impurity, see the r.~h.~s.\ of Fig. \ref{fig2}. 

\subsection{Bond-impurity}\label{secbond}
For the non-interacting system  ($V=0$) with one bond-impurity,
it is easily seen
 that the
back-scattering contributions, i.~e.\ the Fourier components with $q=\pi$ in the
half-filled case, 
cancel each other as is the case for two site-impurities
at an odd distance.
For this reason, we presume that the phase sensitivity 
increases  slightly with 
system size according to $N\Delta E\propto (1-2bN_0/N)$
also in the interacting system.
For repulsive interaction  
the results of Eggert \& Affleck \cite{Eggert92} show that
this kind of  perturbation is irrelevant and the system remains metallic.
For an attractive interaction, the behaviour is expected to be similar.
Accordingly, the results for {\em integrable, triangle like defects}
\cite{Schmi95}
show that 
the phase sensitivity is slightly reduced for non-interacting fermions
in the presence of such (transparent) impurities. In particular,
a repulsive interaction leads to stronger suppression of the phase sensitivity
than an attractive one.

\begin{figure}
{\psfig{figure=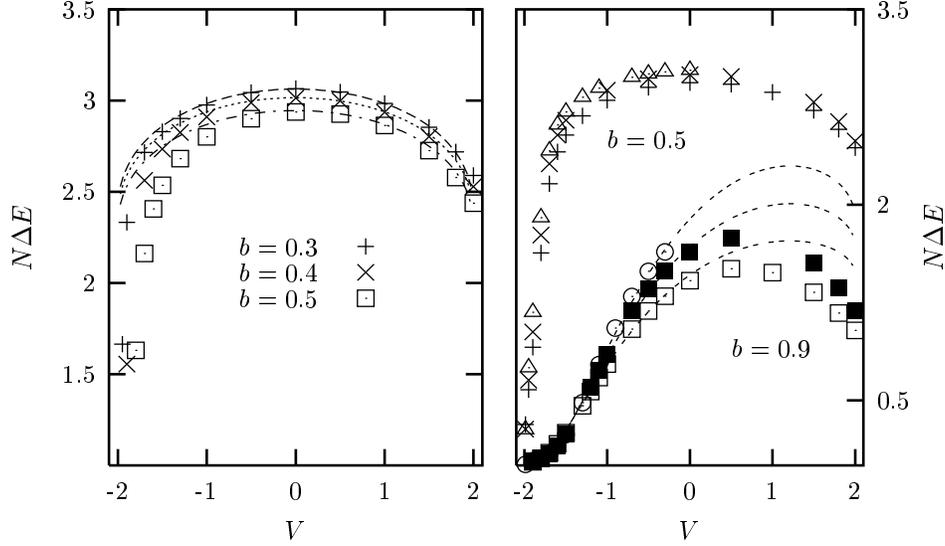,width=1.0\columnwidth}}
  \caption{\label{figb}
Left: Phase sensitivity as a function of interaction for a system with one bond 
impurity of varying strength, 
the system size is $N=40$. 
The lines correspond to $N\Delta E(b=0)(1-2bl_0/N)$.
Right: Phase sensitivity as a function of interaction for a system with one bond
impurity of strength $b=0.5$
($N=40$: $+$, $N=60$: $\times$, $N=80$: $\triangle$),
and $b=0.9$ ($N=40$: $\square$, $N=60$: $\square$,
$N=80$: $\square$).
 The lines correspond to fits according to Eq. \eqref{beff}.}
\end{figure}

The numerical data presented in 
the left part of Fig. \ref{figb} confirm this assertion of an only slight
reduction: clearly 
the system behaves metallic for small $b$. However,
by increasing the impurity strength, strong deviations  are visible,
already for $b=0.3$ at $V=-1.9$. The value of the interaction, where
this strong decrease is seen, moves to a larger interaction 
when increasing the impurity strength, until the maximum reaches the
repulsive regime for $b>0.7$.
Using the scaling applicable for the weak link case, Eq. \eqref{eq15},
$t_t\to t_t^{\rm eff}=t_tN^{1-1/g}$, 
a scaling relation for the two-weak-link case, given by the Hamiltonian in Eq. \eqref{hamb},
can be derived,
\begin{equation}\label{beff}
(1-b)\to(1-b)_{\rm eff}\propto (1-b)N^{1-g/2-1/(4g)}.
\end{equation} 
According to this scaling relation, the weak links increase for $V> -1.2$.
Thus,
for an attractive interaction, there is an interaction value
for which the phase sensitivity becomes
independent of system size, while it scales to zero 
when further lowering the interaction 
down towards $V=-2$ 
(where the first-order transition takes place in the clean system).
Nevertheless, the system remains metallic in this region, but with a strongly
reduced Drude weight.
This behaviour, see the r.~h.~s.\ of Fig. \ref{figb}, is
different from the behaviour of an integrable defect (where a strong reduction
arises for a repulsive interaction).

\section{Combination of dimerization and impurity }\label{sec4}
\subsection{A barrier in a dimerized system}
\begin{figure}
\centerline{\psfig{figure=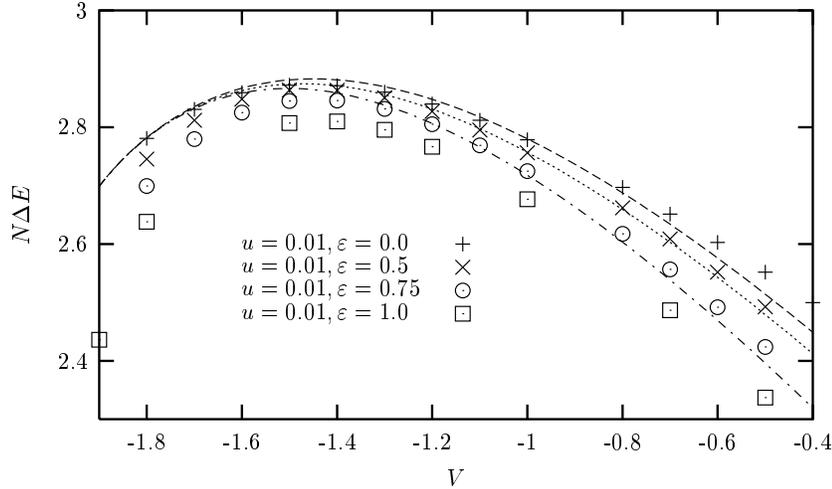,width=\figwidth}}
\caption{\label{uefit}
Phase sensitivity versus interaction for a clean and a distorted
dimerized system;
the system size is 
$N=48$.
The lines correspond to
Eq. (\ref{oned}).}
\end{figure}

We combine now two perturbations and study first a site-impurity in the
dimerized system. For a first impression,
we calculate the
energy levels and the ground-state energy of  
of non-interacting particles with an
alternating hopping, following the treatment in  \cite{SSH}, 
but adding one potential scatterer of strength $\epsilon$.
The ground-state energy is given by
\begin{equation}
E=-2t\sum_{-k_F}^{k_F}\sqrt{\cos^2{ka}+\frac{\epsilon^2}{4N^2}+u^2\sin^2{ka}}
\propto{\rm E}(\kappa)
\end{equation}
with E$(\kappa)$ the second complete elliptic integral, where $\kappa$
is given by $\kappa^2=(1-u^2)/(1+\epsilon^2(N))\sim 1-u^2-\epsilon^2(N)$;
$\epsilon(N)=\epsilon/2N$.
A single impurity in the non-interacting alternating chain
was also discussed in \cite{Segal}.
Thus the phase sensitivity is given by, compare \cite{EPJB}:
\begin{equation}\label{oned}
N\Delta E=N\Delta E(0,0)-2N_0\sqrt{u^2+\epsilon^2(N)}\left(\frac{N}{N_0}\right)^{2-g}.
\end{equation}
Figure \ref{uefit} shows numerical data in comparison to this formula,
with $N_0\approx 4$ according to the cell doubling in the dimerized state.

\begin{figure}
\centerline{\psfig{figure=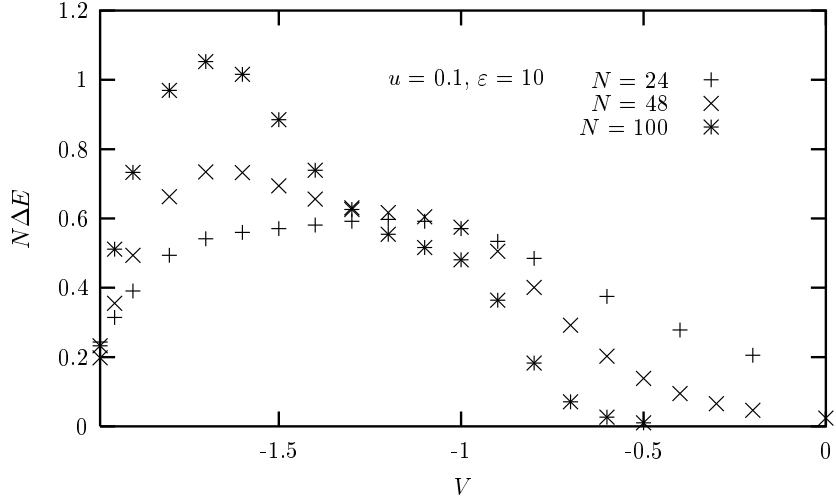,width=\figwidth}}
\caption{\label{u1e10}
Phase sensitivity as a function of interaction for a system with $u=0.1$ and
$\epsilon=10$. For $V>-1$ the phase sensitivity decreases with system size,
and for $V<-1.4$ it increases. In between, it increases from $N=24$ to $N=48$, 
but decreases from $N=48$ to $N=100$.}
\end{figure}

The length dependence presently is not as clear as for a single impurity 
or for the clean dimerized system:
For $g<1$ the system is completely localized, the phase sensitivity 
decreases with system size, while
for $g>2$, the system appears to be delocalized. In between, 
for $ 1<g<2$, however,  a characteristic system size $N^2_c=
\epsilon^2(g-1)/[4u^2(2-g)]$ can be defined. For $N$ smaller than $N_c$,
the phase sensitivity increases with $N$ similar to the behaviour 
in the delocalized phase. Increasing the system size above $N_c$, however,
the phase sensitivity decreases, indicative of localization.
This intermediate regime is shown 
in Fig. \ref{u1e10}, where we plot the
phase sensitivity for a set of parameters which clearly shows
the described behaviour.
\begin{table}[b]
\begin{center}
\begin{tabular}{|c|c||c|c|}
\hline
$u_V=0$ &&& \\
$u$ & $\epsilon$ & $V_{\rm peak}$ & $V_c$ \\
0.1  & 10 &   (-1) & -1.3  \\
0.15 & 10 &  -1.1 & -1.4  \\
0.2 &  10 & -1.2 & -1.6  \\
0.25 & 10 & -1.2 & -1.7  \\
0.3 &  10 & -1.2 & -1.8 \\
\hline\end{tabular}
\begin{tabular}{|c|c||c|c|}
\hline
$u_V=u_t$ &&& \\
$u$ & $\epsilon$ & $V_{\rm peak}$ & $V_c$ \\
0.1  & 10 &   & -0.9   \\
0.2 &  10 &  & -1.3  \\
0.3 &  10 & -1.0 & -1.45 \\
0.5 &  10 & -1.05 & -1.65 \\
0.7 &  10 & -1.1 &-1.75  \\
\hline\end{tabular}
\end{center}
\caption{\label{tab1}Values of $V_{\rm peak}$ and $V_c$ for $u_V=0$, 
and $u_V=u_t$. The latter case is included for completeness, compare
the discussion in subsection 2.2,  
even though not mentioned separately in the text.}
\end{table}

\begin{figure}
\centerline{\psfig{figure=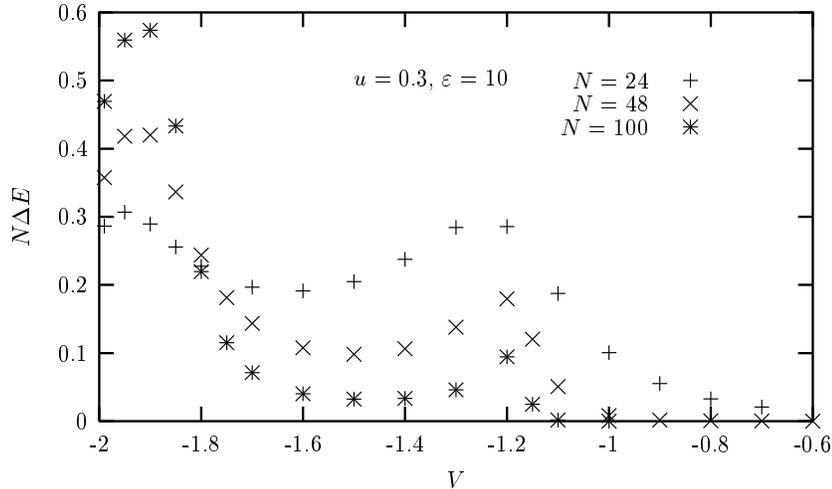,width=\figwidth}}
\caption{\label{u3e10}
Typical plot of the phase sensitivity versus interaction for a system with 
strong dimerization and a strong impurity.}
\end{figure}

An impurity with weak or intermediate strength enlarges the delocalized region
of the dimerized model, as already shown in \cite{Annalen}.
For a strong barrier and a strong dimerization, a new feature is 
observed: In addition to the enlargement of the delocalized region a peak
in the phase sensitivity is found in the localized regime, for 
$\epsilon > 3$ and $ u > 0.1$. The characteristic behaviour
is shown in Fig. \ref{u3e10}, demonstrating that the increase 
for a value of $V_{\rm peak}\approx -1$ is already by one order 
of magnitude for $N=100$.
The assumption that the impurity becomes irrelevant at $V_{\rm peak}$ 
is consistent with the
initial increase of the phase sensitivity at this value.  
Since the effect of the 
dimerization is weakened by an 
impurity, the following (by further lowering the interaction down to 
the value $V_c$)
decrease of the phase sensitivity can be explained by the enhancement
of the effective dimerization,  
caused by the diminished influence of the impurity.
Lowering  the interaction further, the dimerization becomes
irrelevant, too, and the system delocalizes.
Table \ref{tab1} shows the characteristic interaction values for
this model. 
\subsection{Dimerization and bond-impurity}
\begin{figure}
\centerline{\psfig{figure=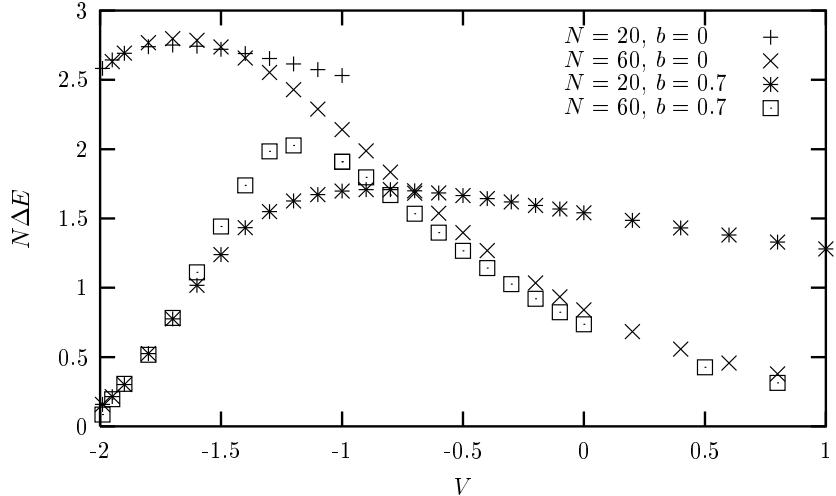,width=\figwidth}}
\caption{\label{ub}
Phase sensitivity as a function of interaction for a dimerized
($u=0.03$) system  with a strong bond-impurity, $b=0.7$;
the system sizes are $N=20$ and $N=60$, respectively.
For comparison, the data of the clean dimerized system, again for 
$N=20$ and $60$,
are included.}
\end{figure}

While we analyzed above
a single site-impurity in a dimerized ring,
we now concentrate on a single bond-impurity plus dimerization.
Based on the results of Sec. \ref{secbond}, we expect that for a weak distortion
the system still shows the phase 
transition at $V=-\sqrt{2}$, with the delocalized region
extending to $V=-2$.
Increasing the strength of the bond-impurity, we also expect the
unusual behaviour -- as described above --
for strong attractive interaction to occur in the dimerized system as well,
because in this interaction regime the dimerization
is irrelevant. 
The first conjecture can be confirmed numerically for strong dimerization
but small $b$.
The numerical data  for stronger distortion, i.~e. 
 for increasing $b$,  show a more complex
behaviour of the phase sensitivity, see Fig. \ref{ub}.
The first observation is, similar to a site-impurity, that the
transition to the metallic state occurs at weaker interaction strength,
e.~g.\ for $u=0.03$ already at $V=-1$ (compared to
$V\approx -1.5$ in the undistorted case). Also, the second 
conjecture is supported by the numerical data: For an interaction
larger than $V\approx -1.5$ the influence of the bond-impurity
drives the systems again to a metallic state with a 
strongly reduced phase sensitivity.
In between, $-1.5<V<-1$,
the system flows to the free metallic fixed point when increasing 
the system size. 
\section{Summary}\label{sec5}
Our numerical studies of a one-dimensional spin chain, which is
equivalent to a system of spinless interacting fermions, 
with various distortions show 
that both types of site-impurities are irrelevant for an attractive 
interaction (using the fermion picture), and
relevant for a repulsive one. The description within a 
first-order perturbation treatment can be applied
in the case of the site-impurity. 
The bond-impurity is irrelevant for repulsive interaction, and leads to a
strong suppression of quantum coherence -- but not to a localized ground-state
for attractive interaction near the first-order transition. 
The scaling could be determined using the bosonization
technique.
In the dimerized system, the numerical results show that
dimerization
and any kind of impurity weaken each other. 
A single  barrier in a dimerized chain leads to an enhancement of
the delocalized region compared to  the
clean dimerized system.
In addition, a sharp peak (as a function of the interaction parameter), 
increasing  the phase sensitivity by
one to two orders of magnitude for small systems, $N\approx 100$,
is found for strong perturbations.

A bond-impurity in the dimerized system is irrelevant 
for a small impurity strength. A stronger impurity leads to
a complicated phase diagram, especially for an attractive interaction.  
The combination of dimerization and impurity therefore leads in all cases
to a shift of the localization-delocalization transition of the dimerized system
to a weaker interaction strength. 
By tuning the values of dimerization and impurity strength, different
metallic windows in the insulating system can be opened. At an interaction
of about $V\approx -0.5, \ldots, -1$ a bond-impurity  
turns the system into the metal. 
For a strong attractive interaction,  $V\approx -1.5, \ldots, -2$, 
where the bond-impurity reduces the Drude weight drastically, a strong 
site-impurity leads to delocalization, especially in the 
strongly dimerized system. 

\vspace*{0.25cm} \baselineskip=10pt{\small \noindent The authors gratefully 
acknowledge financial support by the Deutsche Forschungsgemeinschaft
(through SFB 484 and SPP 1073).  }

\end{document}